\documentclass[aps,prb,showpacs,preprintnumbers,twocolumn,amsmath,amssymb,nobibnotes,superscriptaddress]{revtex4-1}

    \usepackage{xcolor}
    \usepackage{graphicx}
    \usepackage{dcolumn}
    \usepackage{bm}
    \usepackage{graphicx}

\begin{document}

    \title{Ab-initio exploration of Gd monolayer interfaced with WSe2: from electronic and magnetic properties to the anomalous Hall effect}

    \author{Lyes Mesbahi}
    \email[Corresponding author : ]{lyes.mesbahi@ummto.dz}
    \affiliation{Laboratoire de Physique et Chimie Quantique (LPCQ), Mouloud Mammeri University of Tizi-Ouzou, 15000 Tizi-Ouzou, Algeria}
    \author{Omar Messaoudi}
    \affiliation{Laboratoire de Physique et Chimie Quantique (LPCQ), Mouloud Mammeri University of Tizi-Ouzou, 15000 Tizi-Ouzou, Algeria}
    \author{Hamid Bouzar}
    \affiliation{Laboratoire de Physique et Chimie Quantique (LPCQ), Mouloud Mammeri University of Tizi-Ouzou, 15000 Tizi-Ouzou, Algeria}
    \author{Samir Lounis}
    \affiliation{Institute of Physics, Martin-Luther-University Halle-Wittenberg, 06099 Halle (Saale), Germany}
    
  \date{\today}

    \begin{abstract}
        
        Heterostructures involving transition metal dichalcogenides (TMDs) have attracted significant research interest due to the richness and versatility of the underlying physical phenomena. In this work, we investigate a heterostructure consisting of a rare-earth material, specifically a Gd monolayer, interfaced with WSe$_2$. We explore its electronic structure, magnetic properties, and transport behavior, with particular emphasis on the emergence of the anomalous Hall effect (AHE). Both Gd and W are heavy elements, providing strong spin-orbit coupling (SOC), which plays a crucial role in triggering the AHE. The combination of strong SOC and inversion symmetry breaking leads to pronounced asymmetries between the $\Gamma-K$ and $\Gamma-K^\prime$ directions in the Brillouin zone. Our calculations reveal a substantial anomalous Hall conductivity (AHC) at the ferromagnetic interface, primarily originating from numerous avoided crossings involving the d-states of both Gd and W near the Fermi level. Moreover, we demonstrate that the AHC is highly tunable, either by adjusting the in-plane lattice constant or by reducing the separation between Gd and WSe$_2$.

    \end{abstract}

    \maketitle

    \section{Introduction}\label{sec1}

    Two-dimensional (2D) Van der Waals materials are an interesting category of materials, with significant potential for driving innovative technological advancements especially spintronics devices. Among 2D materials, transition metal dichalcogenides (TMDs) have emerged as particularly versatile candidates, offering a wide range of electronic behaviors from semiconducting \cite{abbadi2020ab,mak2010atomically} to metallic \cite{freitas2015,li2014}, semi-metallic \cite{lee2015tungsten}, and even superconducting phases \cite{li2017superconducting,wu2019ion,chen2020strain,mutka1983}. TMDs have the general formula \( \text{MX}_2 \), where \( \text{M} \) is a transition metal atom and \( \text{X} \) is a chalcogen atom (S, Se, or Te). They are classified into different groups based on their transition metal, with each group exhibiting distinct electronic and structural properties. Group IV TMDs, namely \( \text{MoX}_2 \) and \( \text{WX}_2 \) (\( \text{X} = \text{S}, \text{Se}, \text{Te} \)), stands out as some of the most extensively studied systems. The two systems share the same electronic properties, both being semiconductors when X = S and Se. However, W-based systems have the advantage of exhibiting more pronounced effects related to spin-orbit coupling (SOC), one of which is large band splitting \cite{kosmider2013large,liu2020temperature}.
    Group IV TMDs crystallize in a honeycomb-like structure composed of a hexagonal arrangement of atoms, where each transition metal atom is surrounded by six chalcogen atoms in a trigonal prismatic coordination. This geometric arrangement is preserved both in the bulk and in the monolayer forms. Thanks to the van der Waals interactions between the layers, it is relatively easy to isolate monolayers of these materials via mechanical exfoliation \cite{huang2020universal}. However, when reduced to a monolayer, the system loses its inversion symmetry, which, combined with the presence of strong spin-orbit coupling (SOC), leads to the emergence of novel electronic properties. In particular, electrons located at the corners of the Brillouin zone, at  K and  K$^\prime$ , acquire opposite orbital angular momentum, enabling phenomena such as valley polarization and various topological effects \cite{xiao2012coupled,song2015tunable}. Despite their promising features, the lack of intrinsic magnetism in the group IV  TMDs limits their application in certain spintronic devices, especially those relying on the anomalous Hall effect (AHE) or similar phenomena. Recent advances have focused on addressing this limitation by introducing a magnetic proximity effect. This can be achieved by placing a magnetic layer \cite{messaoudi2018nondegenerate} or substrate \cite{habe2017anomalous} in proximity to the TMD, which induces a local magnetic moment and allows for the observation of magnetic effects. Alternatively, the incorporation of magnetic dopants or heterostructures containing magnetic materials \cite{lai2019tunable} can further enhance the spintronic properties of these systems, paving the way for new functionalities in quantum information processing and low-power electronics.
    In this paper, we present the effect of interfacing a Gd layer with a monolayer of WSe$_2$. The study was performed ab-initio using Density Functional Theory (DFT) to calculate the electronic and magnetic properties of the heterostructure. Our calculations predicts the apparition of states within the gap of the WSe$_2$ layer due to hybridization with the ferromagnetic Gd monolayer, making the interface ferromagnetic and metallic with an out of plane magnetic moment of 7.36 $\mu_B$. The bands located around the Fermi level are strongly influenced by the SOC induced by W and Gd. We also observed that the inversion symmetry breaking in the heterostructure Gd/WSe$_2$ translates to non equivalent bands along $\Gamma$-K and $\Gamma$-K$^\prime$,  the so called valley Zeeman splitting \cite{li2014valley,aivazian2015magnetic}, this leads to valley-dependent phenomena. Our study shows that the lattice parameter of the heterostructure is slightly larger than that of the monolayer WSe$_2$, with the difference being 2\%. This prompted us to investigate the effect of varying this parameter on the underlying electronic and transport properties of the interface. Such approaches have already been demonstrated in experimental studies on group-IV monolayers, where uniaxial strain was applied through a four-point bending apparatus \cite{conley2013bandgap} or a cantilever device \cite{he2013experimental}, while biaxial strain was realized via thermally expanded or compressed polypropylene (PP) substrates \cite{frisenda2017biaxial}. We found that the structural effects were minimal, with only slight variations observed, particularly in the distance between the WSe$_2$ monolayer and the Gd monolayer, as well as in the distance between the Gd monolayer and the Se atoms of the WSe$_2$ monolayer. However, the electronic structure was notably affected by changes in the lattice parameter, especially at the Fermi level, where the bands tended to flatten in the K and K$^\prime$ regions. This behavior has significant implications for the Berry curvature in the region and, consequently, for the AHE, a key phenomenon in spintronics.

    \subsection{Computational Details}\label{subsec2}

    Electronic ground-state calculations were performed using DFT with a plane-wave basis, as implemented in the Quantum Espresso 7.3 simulation package \cite{giannozzi2009quantum}. The variable-cell relaxation method was employed to optimize both the unit cell parameters and atomic coordinates in DFT+U, to consider the strongly localized f-electrons of Gd. The PBE-GGA \cite{perdew1996generalized} pseudopotential \cite{heine1970pseudopotential} was used to describe exchange-correlation effects, with a Hubbard parameter \cite{anisimov1991band,shick1999implementation} of $3.5~eV$ acting on the $f$-states calculated using the Hubbard Parameters (HP) module implementing density-functional perturbation theory (DFPT) in Quantum Espresso \cite{timrov2022hp}. The associated electronic and magnetic properties were found to remain unchanged within the range $3$ to $4$ eV. To ensure good convergence, we set the following DFT parameters: an energy cutoff (Ecut) of $200~Ry$, a $30~\mathrm{\AA}$ thick vacuum layer to prevent interactions between periodic repetition, a Monkhorst-Pack \cite{pack1977special} K-mesh of $15\times15\times1$ for self-consistent field (SCF) calculations, and $30\times30\times1$ for non-self-consistent field (NSCF) calculations. The convergence criteria were set to $10^{-8}~Ry $ for energy and $10^{-3}~Ry$ for forces.
    The volume cell relaxation (vc-relax) calculation yielded a lattice parameter of $a = 3.39~\mathrm{\AA}$. In this study, we focused on three distinct lattice parameters: $a = 3.32~\mathrm{\AA}$, corresponding to the lattice parameter of a monolayer WSe$_2$; $a = 3.39~\mathrm{\AA}$, obtained by relaxing both the cell and atomic positions; and $a = 3.44~\mathrm{\AA}$, to assess the influence of increasing the lattice parameter. 

    \begin{figure}[h]
    \centering
    \includegraphics[width=0.5\textwidth]{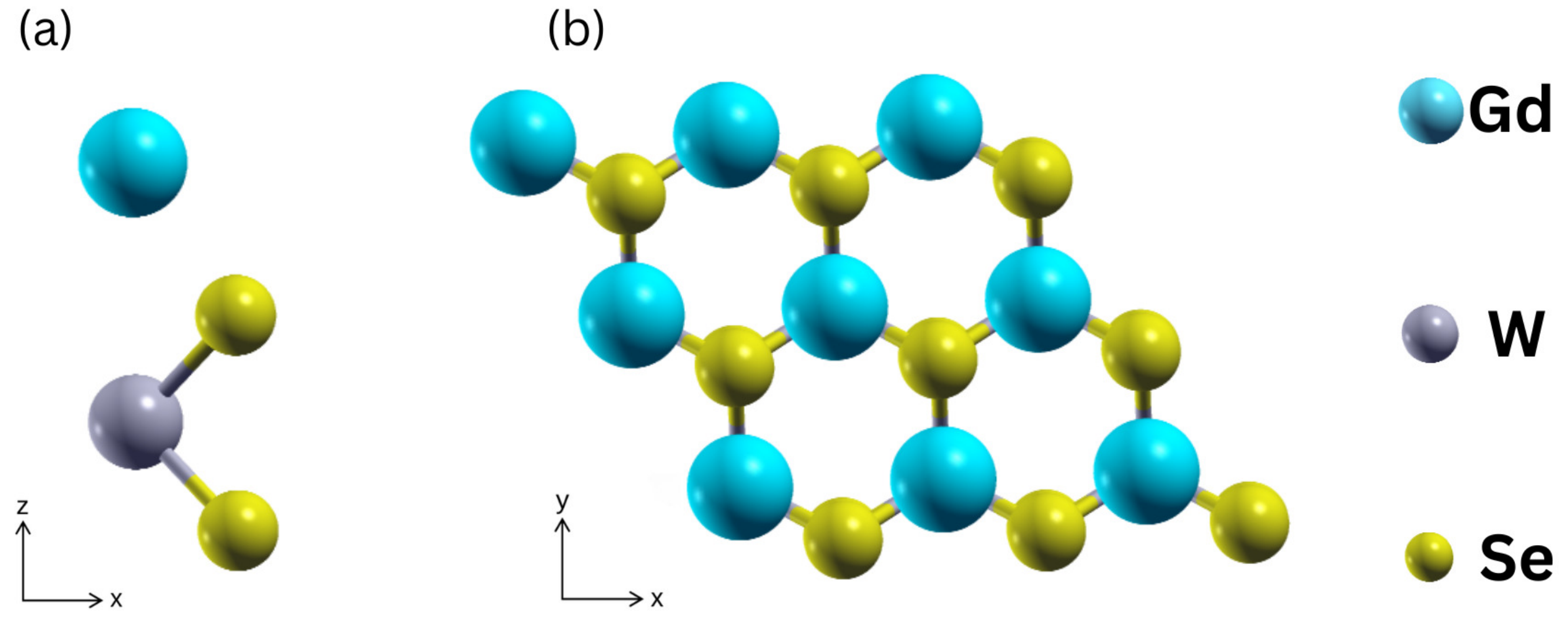}
    \caption{ Crystal structure of Gd/WSe$_{2}$. (a) and (b) respectively show the side and top views.}\label{structure}
    \end{figure}

    \begin{table}[h]
    \begin{tabular}{@{}cccccc@{}}
    \toprule
    $a  (\mathrm{\AA})$ & Gd-W $ (\mathrm{\AA})$  & Se-Se $(\mathrm{\AA})$ & W-Se $(\mathrm{\AA})$ & E(eV) &$m_{S}$ ($\mu_{B}$)\\
    3.32   & 4.13  & 3.42  & 2.60 & 0.0812 & 7.40 \\
    3.39   & 4.05  & 3.38  & 2.61 & 0.0000 & 7.36  \\
    3.44   & 3.93  & 3.32  & 2.61 & 0.0138 & 7.31  \\
    \botrule
    \end{tabular}

    \caption{Interatomic distances, total energy (with the energy of the relaxed structure as reference) and magnetic moment as a function of the lattice parameter of the Gd/WSe$_2$ heterostructure.} \label{table_param}%
    \end{table}

    Table 1 shows that the monolayer structure is barely altered, as the increase in the a parameter is compensated by the reduction of the distance between the two selenium atoms. Additionally, the distance between the Gd atom and the W atom decreases in correlation with the increase of the lattice parameter. For the energy difference, we took the energy of the vc-relax structure as a reference, and our calculations show that the system is significantly more stable with this lattice parameter, with an energy difference on the order of $10^{-2}~$eV. As for the magnetic moment, it does not vary significantly with the changes in the lattice parameter, the calculated values (Table \ref{table_param}) are in agreement with the literature\cite{carbone2025magnetic}. The magnetic moment of Gd ($7.36~\mu_B$) is predominantly contributed by the f-electrons ($6.95~\mu_B$), with the remaining part arising mainly from the d-electrons. Spin-orbit coupling does not significantly influence the magnitude of the moment. However, coupling to WSe$_2$ leads to a reduction of about $0.24~\mu_B$ compared to the unsupported Gd monolayer, primarily due to a decrease in the d-electron contribution caused by hybridization of the electronic states at the interface.

A previous study has shown that magnetic anisotropy energy was found to reach $0.4 $ meV \cite{carbone2025magnetic}, thus imposing an out-of-plane magnetic moment. The anisotropy was attributed to the strong interaction between the spin-polarized Gd $d$-states and the surrounding $C_{3v}$ crystal field, with both Gd and the spin-polarized substrate contributing similarly to the MAE and favoring an out-of-plane orientation.

    We computed the AHC of the system using the post-processing code provided in the Wannier90 suite \cite{wang2006ab,mostofi2014updated}. We used 36 Maximally Localized Wannier Functions (MLWFs) with a disentanglement window of $13~$eV around the Fermi level to ensure a better convergence. 

    \section{Results}\label{sec2}

    \subsection{Ground states properties}\label{subsec2}
    \begin{figure}
    \includegraphics[width=0.45\textwidth]{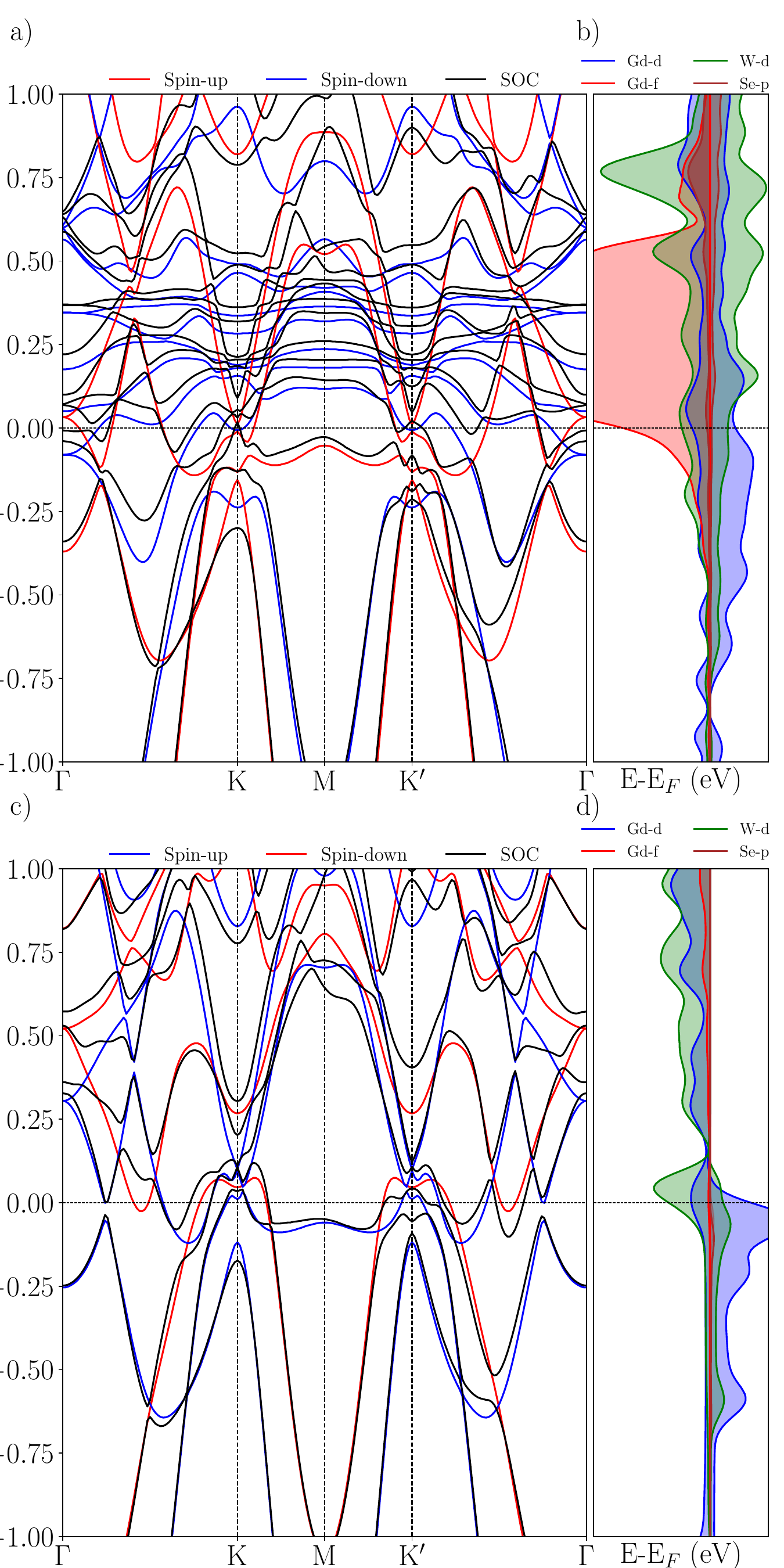}

    \caption{(color online) Calculated electronic structure for Gd/WSe$_2$. (a),(c) Calculated band structures both in the scalar relativistic and fully relativistic cases, for $a = 3.43~\text{\AA}$, U$= 0~\text{eV}$ and  $a = 3.39~\text{\AA}$, U$= 3.5~\text{eV}$, respectively. The red and blue lines represent respectively the spin up and spin down of the scalar relativistic calculation, while the black line represents the fully relativistic bands. (b),(d) Calculated scalar projected density of states} 

    \label{fig2}
    \end{figure}

    In this section, we will first focus on the electronic structure of our system by comparing the results obtained with and without the Hubbard U potential. Subsequently, we will analyze the effect of varying the lattice parameter $a$ on the band structures. Under the effect of the Hubbard U potential, the lattice parameter a changes after the vc-relax calculation; we find $a = 3.39~\mathrm{\AA}$ for the relaxation with U and $a = 3.43~\mathrm{\AA}$ for the relaxation without U, which represent differences of $1.86~\%$ and $3.25~\%$, respectively, compared to the lattice parameter of the isolated WSe$_2$ monolayer. Figure~\ref{fig2} shows the obtained band structures and their corresponding projected density of states. Fig~\ref{fig2}.(a) shows the calculated band structure for U$=0$ and $a = 3.43~\mathrm{\AA}$. The calculated bands for U$=3.5$ eV and the corresponding vc-relax parameter $a=3.39~\mathrm{\AA}$ is shown in Fig~\ref{fig2}.(c).

    We observe that under the influence of the Gd monolayer, our system transitions into metallic phase, as confirmed by the projected density of states (PDOS) with a magnetic moment of $ 7.11~\mu_{B}$ for the calculation without U and $ 7.36~\mu_{B}$. For the relaxed lattice parameter ($a = 3.39~\mathrm{\AA}$), the ferromagnetic state is found to be more stable than the antiferromagnetic one, favored by an energy difference of $55$ meV per Gd atom. As shown in the projected density of states in Fig~\ref{fig2}.(d)  the Gd 5$d$ electrons provide substantial weight at the Fermi level, while the  4$f$ states localized around $-7.5$ eV, well below the Fermi level. Consequently, the stabilization of the ferromagnetic ground state can be understood in terms of an indirect RKKY-type exchange, where the localized  4$f$ moments couple via the more delocalized  5$d$ conduction electrons, consistent with previous reports \cite{kurz2002magnetism,szade1990rkky}.

    In the case where the Hubbard potential is applied, it is apparent that the  $d$-electrons from the Gd atoms are dominant in the Fermi level region, closely followed by the W d-electrons. Without the U potential, the contribution from the Gd $f$-electrons is no longer negligible in this energy region. The contribution of selenium is relatively minimal around the Fermi level. Consequently, we can conclude that the Fermi electrons primarily originate from the hybridization between W and Gd atoms. This behavior is similar to what was found in the Eu/WSe$_2$ study \cite{carbone2025magnetic}, although the most notable distinction lies in the fact that, in our systems, the Gd  $f$-electrons are deeper than those of $f$-Eu ($-7.5~$eV with U, $-5~$eV without U for  $f$-Gd and $-3~$eV for  $f$-Eu. These findings are in line with those from the GdSe$_{2}$ study, where the $f$-electrons are located in a similar energy region as in our systems (in the conduction band and around $-6 ~$eV)\cite{zeer2024promoting}. 
    \begin{figure}[h]
    \centering
    \includegraphics[width=0.45\textwidth]{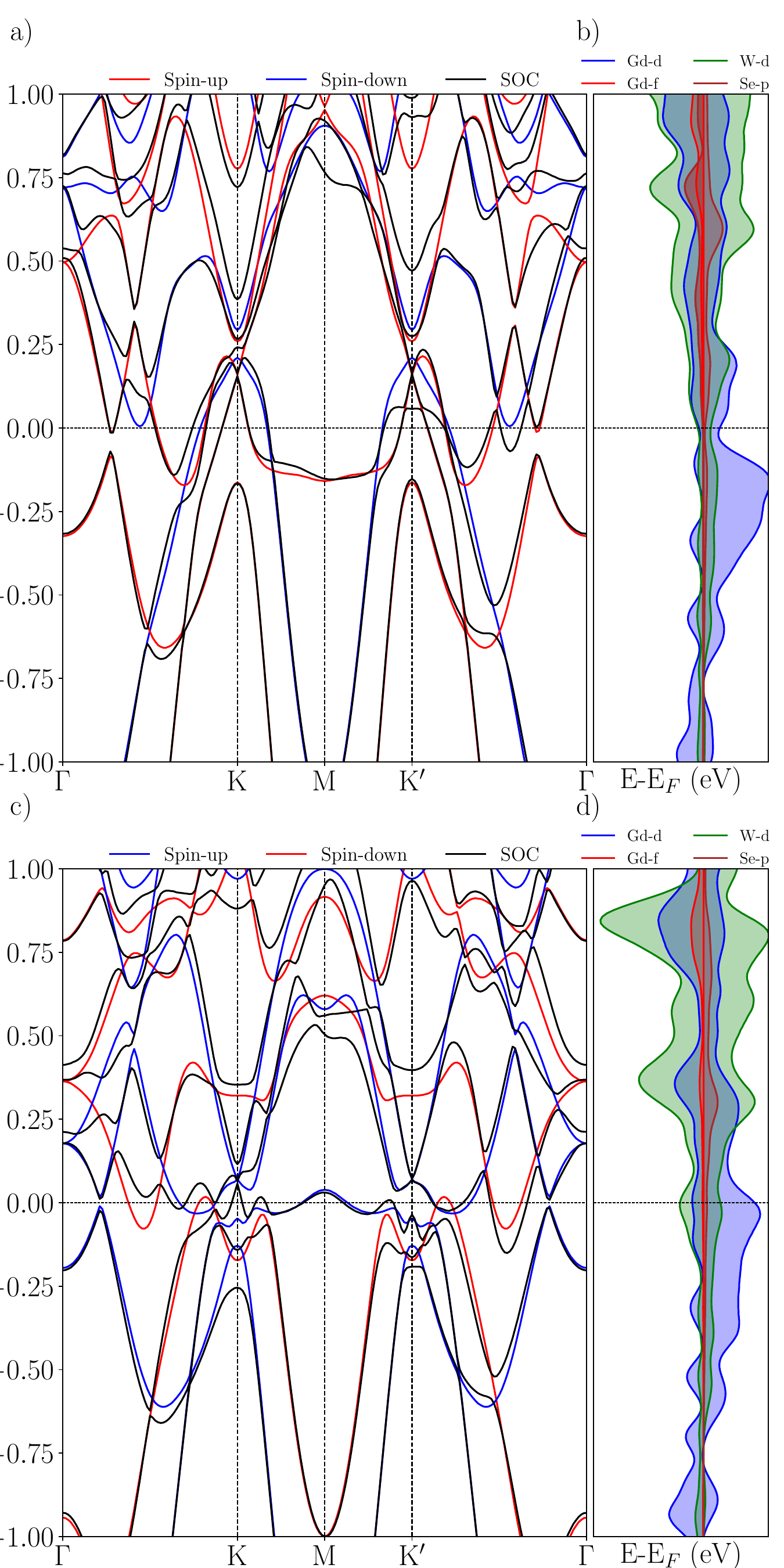}
    \caption{(color online) Calculated electronic structure, for Gd/WSe$_2$. (a),(c) Calculated Band structure assuming a U of $3.548~$eV  both in the scalar relativistic and fully relativistic cases for respectively $a=3.32~\mathrm{\AA}$ and $a=3.44~\mathrm{\AA}$. The red and blue lines represent respectively the spin up and spin down of the scalar relativistic calculation, while the black line represents the fully relativistic bands. (b),(d) Calculated scalar projected density of states}
    \label{fig3}
    \end{figure} 
    Additionally, the $d$-Gd electrons remain dominant around $-2~$eV, while in the Eu/WSe$_2$ study, the Eu-$d$ electrons are displaced by the W-$d$ electrons. The effect of spin-orbit coupling (SOC) in our systems primarily arises from W, as it has an atomic number Z = $74$, whereas Gd has Z = $64$. This effect becomes evident in the band structures, where the splitting is most pronounced in regions where Gd-$d$ and  W-$d$ electrons are present ($-2~$eV) and in the conduction region ($0.5-1~$eV), while it is almost absent in areas where these $d$-electrons are not present ($-1~$to$~-0.6~ ~$eV). Moreover, we observe a notable discrepancy in the influence of SOC along the $\Gamma$-K and $\Gamma$-K$^\prime$ paths, with the difference being particularly apparent halfway through ($\Gamma$-K, $\Gamma$-K$^\prime$) and at both the K and K$^\prime$ points.
    If we now turn our attention to the impact of the lattice parameter variation as shown in Figure~\ref{fig3}, we notice that it mainly influences the region around the Fermi level. The most significant changes in the electronic structure are observed at the K and K$^\prime$ points, where the lobes above the Fermi level (Figure~\ref{fig3} (a)) flatten with the increase of the lattice parameter (Figure~\ref{fig2} (c), Figure~\ref{fig3} (c)). This effect is especially pronounced in the conduction region, where the bands around the K and K$^\prime$ points exhibit noticeable changes. Additionally, at the $\Gamma$ point, we observe a positive energy shift, and the descending lobe near the high-symmetry $M$ point becomes less deep as the lattice parameter increases. The variation in the lattice parameter also impacts the splitting caused by SOC. Specifically, the degeneracy lifting induced by SOC is more evident around the K and K$^\prime$ points, which plays a key role in modulating the Berry curvature. This variation, in turn, significantly influences the AHE.

    \subsection{Anomalous Hall Conductivity}\label{subsec2}
    The calculation of the AHC was performed using the Wannier90 code. The AHC is given by the integral of the Berry curvature over the Brillouin zone, a key quantity in systems where inversion symmetry is broken, such as in ferromagnetic materials. The Berry curvature reflects the variation in the phase of the Bloch wavefunction in k-space and plays a central role in generating the anomalous Hall effect. Notably, the Berry curvature receives significant contributions from regions where the energy bands are separated due to SOC, such as in avoided crossings. These avoided crossings occur when the energy bands approach each other but do not cross, resulting in a gap that affects the Berry curvature and, in turn, the AHC. Previous studies have demonstrated other complex interplay between symmetry and non linear response that could affect the AHC  \cite{nwad114,wang2024orbital}.
    The expression for the anomalous Hall conductivity (\(\sigma^{\text{AH}}\)) is:

    \begin{align*}
    \sigma^{\text{AH}}_{\alpha \beta}(0) &= \frac{e^2}{\hbar} \frac{1}{N_k \Omega_c} \sum_{\mathbf{k}} (-1) \Omega_{\alpha \beta}(\mathbf{k}) \\
    \Omega_{\alpha \beta}(\mathbf{k}) &= \sum_n f_{n\mathbf{k}} \, \Omega_{n,\alpha \beta}(\mathbf{k}).
    \end{align*}

\begin{figure}[h]
    \centering
    \includegraphics[width=0.45\textwidth]{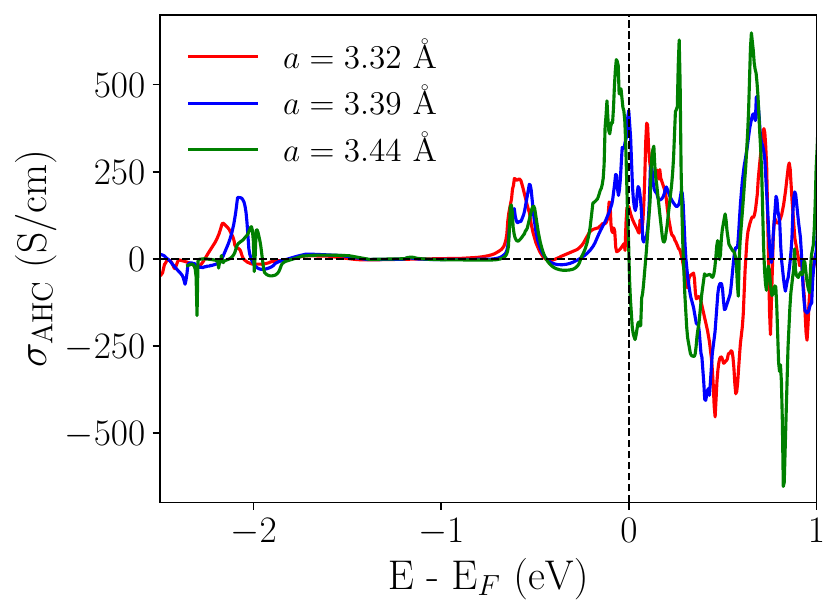}
    \caption{ Anomalous Hall conductivity as function of the Fermi level for $a=3.32~\mathrm{\AA}$, $a=3.39~\mathrm{\AA}$, $a=3.44~\mathrm{\AA}$ respectivly in red, blue and green lines.}
    \label{fig4}
    \end{figure}

    In this equation, \(\Omega_{\alpha \beta}(\mathbf{k})\) represents the Berry curvature, calculated at each \(\mathbf{k}\)-point in the Brillouin zone. The term \(f_{n\mathbf{k}}\) is the Fermi-Dirac distribution of electrons. The cell volume \(\Omega_c\) and the number of points \(N_k\) used to mesh the Brillouin zone are also essential for the accuracy of the calculation.
    For the AHC calculation, the Fermi energy is adjusted to vary the occupation of the electronic bands without altering the shape of the band structure, this allows us to investigate how changes in band occupation influence the AHC.
    The results for the AHE in our system are shown in Figure~\ref{fig4}, where the evolution of the anomalous Hall conductivity is depicted as a function of lattice parameter and band occupation. 

    \
    

    The three curves shown in the Figure~\ref{fig4} present the same features overall, where peaks emerge in the AHC due to band-splitting induced by the interplay of spin-orbit coupling and inversion symmetry breaking, leading to avoided crossings\cite{nagaosa2010anomalous} and inequivalent band-shifts due to the change of the Berry curvature's sign \cite{feng2016,berrysign1,berrysign2}. By changing the lattice parameter, both the intralayer and interlayer distances affect electronic structure hybridization, which in turn impacts the position and shape of the bands, ultimately affecting the magnitude of the AHC. The latter is found to be significant around $-0.6~$eV and experiences an impressive increase around the Fermi level (Figure \ref{fig4}). For instance, at $-0.6~$eV, band-splittings occur between $\Gamma$-K and $\Gamma$-K$^\prime$ (see Figures ~\ref{fig2} and~\ref{fig3}).
    There is a slight variation in the amplitude of the avoided crossings and surrounding curvature, impacting the Berry curvature, which leads to the splitting of the peak observed for a lattice parameter $3.32~\mathrm{\AA}$ (compared to what is found for $a = 3.39$ and $3.44 ~\mathrm{\AA}$). 
    In the neighbourhood of the Fermi energy, more band-splittings appear in a wider area of the Brillouin zone as the lattice parameter increases, see Figures ~\ref{fig2} and~\ref{fig3}. The AHC at the Fermi level changes from $150 ~\text{S/cm}$ for a lattice parameter of $3.32~\mathrm{\AA}$, to $426~\text{S/cm}$ for the relaxed lattice parameter ($3.39 ~\mathrm{\AA}$). While it goes down to $-54~\text{S/cm}$ for \(a=3.44~\mathrm{\AA}\) at the Fermi level, the immediate surroundings of this region exhibit a maximum conductivity of $567~\text{S/cm}$ at $-0.06$ eV below the Fermi level, and $-225~\text{S/cm}$ at $0.03$ eV above. This can be explained by notable changes in the band structure around the K and K$^\prime$ with respect to the lattice parameter increase.
    Consequently, the bands that were initially above the Fermi level at the K and K$^\prime$ points for $a = 3.32~\mathrm{\AA}$ shift down in energy, moving closer to the Fermi level or slightly below it for $a = 3.44~\mathrm{\AA}$, leading to an increased number of band crossings in this region. 

    \ 

    Analyzing the nature of the bands involved, we find that the dominant orbitals in these regions remain unchanged when the lattice parameter increase. The band crossings around $-0.6~\text{eV}$ mainly involve the W $d_{x^2-y^2}$ and $d_{xy}$ orbitals, with a weaker contribution from $d_{zx}$ and $d_{zy}$, all of which are coupled to the Gd $d_{x^2-y^2}$, $d_{xy}$, $d_{zx}$, and $d_{zy}$ orbitals. In addition, the electronic states near the Fermi level at the $K$ and $K'$ points are largely dominated by the Gd-$d$ orbitals ($d_{zx}$, $d_{zy}$, and $d_{x^2-y^2}$), which couple to the W $d_{zx}$, $d_{zy}$, and $d_{z^2}$. Around the $\Gamma$ point, a strong contribution from the Gd-$d_{z^2}$ orbital is observed.

    \begin{figure}[h]
    \includegraphics[width=0.45\textwidth]{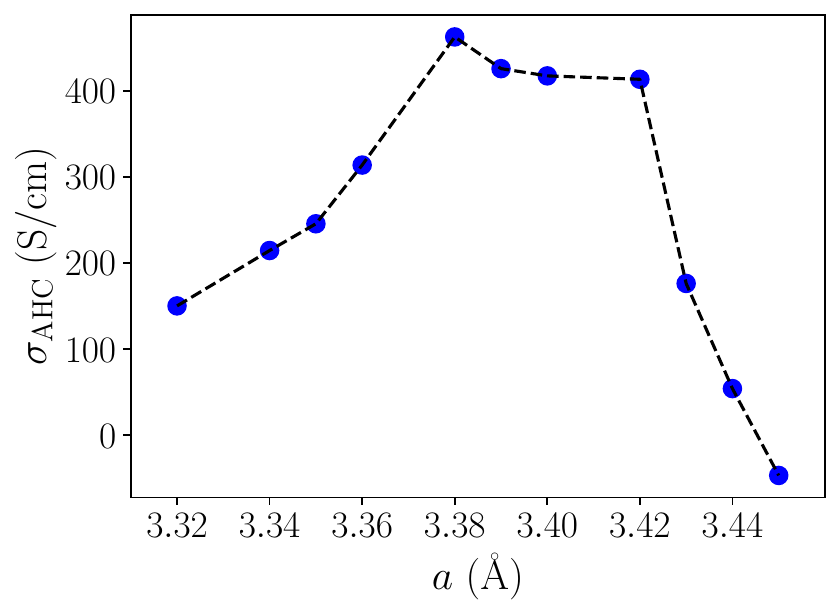}
    \caption{Anomalous Hall conductivity at the Fermi level as a function of the lattice parameter.}
    \label{fig5}
    \end{figure}

    In Figure~\ref{fig5}, we provide an overview by considering only the AHC exactly at the Fermi point, while considering more points for the lattice parameter. We observe an initial increasing trend of the AHC, reaching a maximum of $463~\text{S/cm}$ at $a=3.38~\mathrm{\AA}$, translating into a dramatic 209\% increase in the AHC for a moderate lattice expansion of 1.8\%, followed by a plateau until $a=3.42~\mathrm{\AA}$, and then a decrease as the lattice parameter increases further.
    We compared our results against other studies focusing on the AHC in 2D materials.  SrRuO$_{3}$, for instance, exhibits a peak AHC of $-157 ~\text{S/cm}$ under a $0.47~\%$ strain \cite{samanta2021tailoring}. Similarly, the AHC at the Fermi level in Mn$_{2}$NO$_{2}$, Mn$_{2}$NF$_{2}$, and Mn$_{2}$N(OH)$_{2}$, with respective values of $470~\text{S/cm}$, $66 ~\text{S/cm}$, and $147~$ S/cm \cite{jena2022surface} are comparable to our results ($426 ~$S/cm, $-54~ $S/cm and $150~ $S/cm). We also compared our results to those obtained for the WSe$_2$/VSe$_2$ bilayer, where the AHC values range from $40~\text{S/cm}$ to $100 ~\text{S/cm}$ under a perpendicular electric field applied to the surface\cite{marfoua2022reversal}. To enable direct comparison with other theoretical works, we express our results in \( e^2/h \) instead of \text{S/cm} by considering \( 1\,e^2/h \approx 3.874 \times 10^{-5} \,\text{S} \) and our out-of-plane cell length of $30~\mathrm{\AA}$ to eliminate the cm unit. The converted values are expressed in the following table: 

    \begin{table}[h]
    \centering
    \small
    \begin{tabular}{p{1.5 cm} >{\centering\arraybackslash}p{1.5cm} > {\centering\arraybackslash}p{1.5cm} > {\centering\arraybackslash}p{1.5cm} >{\centering\arraybackslash}p{1.5cm}}
    \toprule
    $a\ (\mathrm{\AA}$)& 3.32 &3.38 & 3.39 & 3.44 \\
    S/cm & 150 &463 & 426 & -54 \\
    e$^2$/h & 0.290 & 0.896 & 0.825 & -0.105 \\
    \botrule
    \end{tabular}
    \caption{Converted units from S/cm to $e^2/h$ for the AHC. }
    \label{tab1}
    \end{table}

    The value obtained for the vc-relax configuration, $-0.825\,e^2/h$, matches the AHC found at the Fermi level in Eu/WSe$_2$ \cite{carbone2022engineering}. Furthermore, our results lie within the same order of magnitude as those reported for the ferromagnetic compound FeGeTe$_2$, where the AHC ranges between ($0.4-1.5~e^2/h$)\cite{guo2023modulating}.

    \subsection{Effect of Interface Spacing on AHC}
    In this section, we simulate the effect of pressure along the z-direction, assuming that this pressure does not alter the Bravais lattice of the system. To do so, we manually fix the distance between the W and the Gd atoms before performing our ab-initio calculations (electronic structure, AHC), using the relaxed lattice parameter  $a = 3.39~\text{\AA}$.

    \begin{figure}[h]
    \includegraphics[width=0.45 \textwidth]{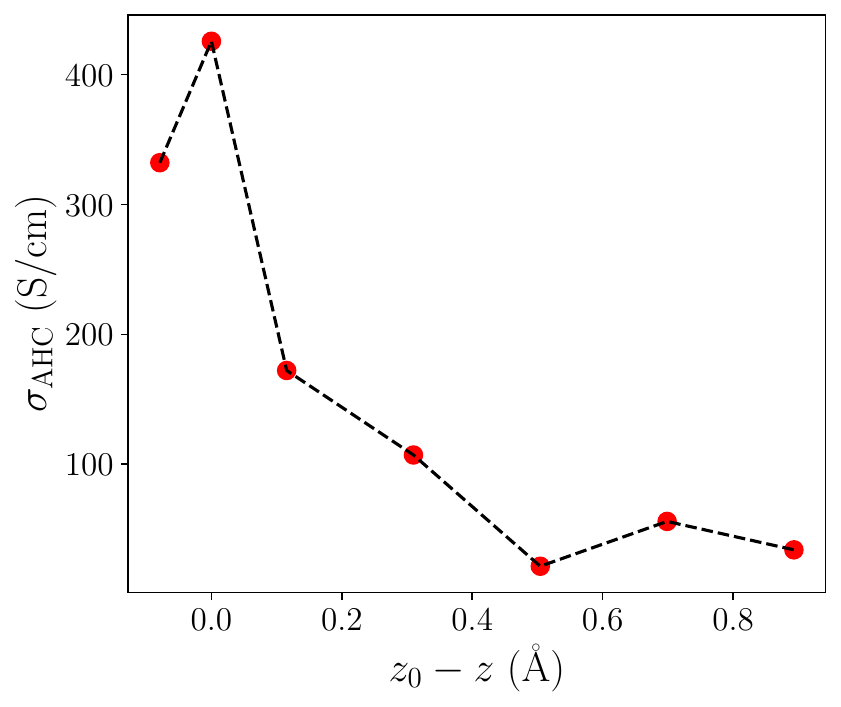}
    \caption{ Anomalous Hall conductivity at the Fermi level as a function of the difference between the distance of the Gd and the W in the relaxed structure $z_0$ and the one in the constrained $z$.}
    \label{fig6}
    \end{figure}
    Just as with the variation of the lattice parameter, the notable changes in the band structure are only visible near the Fermi level. We observe the progressive disappearance of the various band crossings along the $\Gamma$–K and $\Gamma$–K$^\prime$ paths as the distance between the Gd layer and the monolayer is reduced. Furthermore, the bands tend to flatten around the K and K$^\prime$ points.
    The effects of these changes on the AHC are shown in Figure~\ref{fig6}, where $z_0$ is the distance between  Gd and W in the relaxed structure, and $z$ is the distance used in the "constrained" calculations. We observe that the AHC is maximal for $z_0$ and that it decreases drastically as this distance is reduced before stabilizing around $50~\text{S/cm}$. This can be explained by all the changes induced in the electronic bands as a result of the compression of the structure.

    \section{Conclusion}

    Our calculations show that the Gd/WSe$_2$ heterostructure exhibits a large anomalous Hall conductivity (AHC) of $426~\text{S/cm}$, primarily arising from multiple avoided band crossings induced by strong spin-orbit interaction. The system is ferromagnetic, with a large magnetic moment of $7.36~\mu_\mathrm{B}$, and presents a particularly tunable AHC.
    Our work shows that small changes in the lattice parameter (on the order of 2\%) can significantly affect the AHC, reducing it from its initial value of $426~\text{S/cm}$ to about $-50~\text{S/cm}$ under tensile strain and to $150~\text{S/cm}$ under compressive strain, with a maximum value of $462~\text{S/cm}$ observed at $a = 3.38~\mathrm{\AA}$.
    A comparable effect is observed when pressure is applied along the out-of-plane ($z$) direction, bringing the Gd layer closer to the WSe$_2$ monolayer. In this case, we observe a pronounced decrease in the AHC, reaching as low as $47~\text{S/cm}$. These results demonstrate a strong sensitivity of the topological transport properties to both in-plane and out-of-plane structural modifications. The ability to control the AHC over an order of magnitude opens up new possibilities for designing functional spintronic devices based on 2D magnetic heterostructures.

    \bibliography{biblio_ver1}
    \end{document}